\documentclass[12pt,preprint]{aastex}

\usepackage{emulateapj5}

\newcommand{\R}		      {\hbox{\cal R}}
\newcommand{\B}               {\hbox{\cal B}}
\newcommand{\ug}               {\hbox{$u-g$}}
\newcommand{\gr}               {\hbox{$g-r$}}
\newcommand{\ri}               {\hbox{$r-i$}}
\newcommand{\iz}               {\hbox{$i-z$}}
\newcommand{\Po}               {\hbox{$P_1$}}
\newcommand{\Pt}               {\hbox{$P_2$}}

\shorttitle{Metal-poor Giants in the SDSS}
\shortauthors{A. Helmi {\em et al.}}

\begin{document}

\title{Selection of Metal-poor Giant Stars Using the Sloan Digital Sky
Survey Photometric System}

\author{
Amina Helmi\altaffilmark{1},
\v{Z}eljko Ivezi\'{c}\altaffilmark{2,3},
Francisco Prada\altaffilmark{4,5},
Laura Pentericci\altaffilmark{4},
Constance M.~Rockosi\altaffilmark{6}, 
Donald P. Schneider\altaffilmark{7},
Eva K.~Grebel\altaffilmark{4},
Daniel Harbeck\altaffilmark{4},
Robert H.~Lupton\altaffilmark{2},
James E.~Gunn\altaffilmark{2},
Gillian R.~Knapp\altaffilmark{2},
Michael A.~Strauss\altaffilmark{2},
Jonathan Brinkmann\altaffilmark{8}
}

\begin{abstract}
We present a method for photometric selection of metal-poor halo
giants from the imaging data of the Sloan Digital Sky Survey
(SDSS). These stars are offset from the stellar locus in the \gr\
vs. \ug\ color-color diagram. Based on a sample of 29 candidates for
which spectra were taken, we derive a selection efficiency of the
order of 50\%, for stars brighter than $r \sim 17^m$.
The candidates selected in 400 deg$^2$ of sky from the SDSS Early Data
Release trace the known halo structures (tidal streams from the
Sagittarius dwarf galaxy, the Draco dwarf spheroidal galaxy),
indicating that such a color-selected sample can be used to study the
halo structure even without spectroscopic information. This method,
and supplemental techniques for selecting halo stars, such as RR Lyrae stars
and other blue horizontal branch stars, can produce an unprecedented
three-dimensional map of the Galactic halo based on the SDSS imaging
survey.
\end{abstract}


\altaffiltext{1}{Max Planck Institut f\"ur Astrophysik,
85741 Garching bei M\"unchen, Germany. Present address: Sterrekundig
Institut, Universiteit Utrecht, P.O.Box 80000, 3508 TA Utrecht, The
Netherlands. E-mail: ahelmi@phys.uu.nl
\label{MPA}}
\altaffiltext{2}{Princeton University Observatory, Princeton, NJ
08544, United States. E-mail:
ivezic,rhl,jeg,gk,strauss@astro.princeton.edu
\label{Princeton}}
\altaffiltext{3}{Russell Fellow
\label{HNR}}
\altaffiltext{4}{Max Planck Institut f\"ur Astronomie, K\"onigstuhl
17, 69117 Heidelberg,
Germany. E-mail: laura,grebel,harbeck@mpia-hd.mpg.de
\label{MPiA}}
\altaffiltext{5}{Centro Astron\'omico Hispano-Alem\'an, Apartado 511,
E-04080 Almeria, Spain. E-mail: prada@caha.es
\label{Caha}}
\altaffiltext{6}{University of Washington, Dept. of Astronomy, Box
351580, Seattle, WA 98195, United States. E-mail:
cmr@astro.washington.edu
\label{Washington}}
\altaffiltext{7}{Department of Astronomy and Astrophysics, 525 Davey
Laboratory, Pennsylvania State University, University Park, PA 16802,
United States. E-mail: dps@astro.psu.edu
\label{unomas}}
\altaffiltext{8}{Apache Point Observatory, P.O. Box 59,
Sunspot, NM 88349, United States. E-mail: jb@apo.nmsu.edu
\label{otromas}}

\section{Introduction}

Studies of the Galactic halo can help constrain the formation history
of the Milky Way. Currently popular hierarchical models of galaxy
formation predict the presence of substructures (tidal tails, streams)
due to the mergers and accretion the Galaxy may have experienced over
its lifetime (Helmi {\em et al.}~2002; Steinmetz \& Navarro
2002). These structures should be ubiquitous in the outer halo, where
the dynamical timescales are sufficiently long for them to remain
spatially coherent (Johnston {\em et al.}~1996; Mayer {\em et
al.}~2002).  The best tracers of the outer halo are the luminous giant
stars (which can be detected at distances of over 100 kpc), and
several investigations are taking advantage of tailored techniques,
like Washington photometry (Geisler 1984), to identify these stars
(e.g. Majewski {\em et al.}~2000; the Spaghetti Photometric Survey (SPS),
see Morrison {\em et al.}~2000).  In previous works blue horizontal
branch and RR Lyrae stars also have been used to probe the outer halo
(Sommer-Larsen {\em et al.}~1994; Kinman {\em et al.}~1994), and even
discover substructures, many of which are debris from the Sagittarius
dwarf galaxy (Ivezi\'{c} {\em et al.}  2000; Yanny {\em et al.}~2000;
Vivas {\em et al.}~2001).

The Sloan Digital Sky Survey (SDSS; York {\em et al.}~2000) has the
potential of revolutionizing studies of the Galactic halo because it
will provide homogeneous and deep ($r < 22.5$) photometry in five
passbands ($u$, $g$, $r$, $i$, and $z$, Fukugita {\em et al.}~1996;
Gunn {\em et al.}~1998; Smith {\em et al.}~2002; Hogg {\em et al.}
2002) accurate to a few percent of up to 10,000 deg$^2$ in the
Northern Galactic Cap. The survey sky coverage will result in
photometric measurements for about 50 million stars and a similar
number of galaxies. Astrometric positions are accurate to better than
0.1 arcsec per coordinate (rms) for sources brighter than 20.5$^m$
(Pier {\em et al.}~2002). Such a large database is well suited for
studies of Galactic structure, with the caveat that the photometric
system must be able to identify different classes of stars, and
variations in their metallicity and luminosity. This separation is
particularly challenging for Galactic halo tracers, because the number
of halo stars at a given magnitude is much smaller than that of any
other Galactic component (e.g. in a high-latitude field at
$b=45^\circ$, at $V\sim 17^m$ and $0.3 \le (B-V) \le 1.5$ the fraction
of halo giants is $\sim 6$\%, cf. Robin {\em et al.}~2000).

In this paper, we present a method designed to select candidate
metal-poor giants based on their SDSS colors. Using a sample of known
metal-poor halo giants discovered by the Spaghetti Survey, we isolated
a region in the SDSS \gr\ vs. \ug\ color-color diagram where the
probability of a star being a giant is enhanced. Spectroscopic
observations of an unbiased sample of 29 ``candidate" stars with $r <
17$ indicate that the selection efficiency of our technique is
approximately 50\%. In Section \S2 we describe the selection method,
and in Section \S3 we discuss its implications for the Galactic halo
studies.

\section{       Color Selection of Metal-poor Giant Stars                }

\subsection{          The SDSS Photometric Data }

We use SDSS imaging data that were taken during the commissioning
phase, and which are part of the SDSS Early Data Release (Stoughton
{\em et al.}~2002, hereafter EDR). We analyze here equatorial
observing runs 94, 125, 752 and 756 which include two regions:
$|\,\delta_{2000}| \la 1.27^\circ$, and $\alpha_{2000} =$
\hbox{23$^h$ 24$^m$} -- \hbox{03$^h$ 44$^m$} (runs 94 and 125), and 
$\alpha_{2000} = $\hbox{08$^h$ 7$^m$} -- \hbox{16$^h$ 40$^m$} (runs
752 and 756), and cover 394 deg$^2$. In order to test the photometric
repeatability, we also use run 1755 which overlaps with run 125 in the
range $\alpha_{2000} =$ \hbox{23$^h$ 22$^m$} -- \hbox{03$^h$ 03$^m$}
(74.8 deg$^2$ area). Note that the wide range of Galactic coordinates
implies that the different Galactic components will manifest
themselves with varying strength within this dataset.

We have extracted 2,143,248 objects classified as point sources by the
photometric pipeline ({\em photo}, Lupton {\em et al.}~2001),
which do not have any of the following flags set: {\tt bright, satur,
blended, notchecked, deblended\_as\_moving}.  This flag combination 
selects sources with the
most reliable photometry (for details see EDR and Ivezi\'{c} {\em et
al.}~in prep., hereafter Paper II).  We use the ``point-spread
function" magnitudes corrected for interstellar reddening\footnote{The
full reddening correction is applied because the majority of stars
relevant to this study (i.e. blue stars) are expected to be further
than 1 kpc, see Finlator {\em et al.}~(2000).} (Schlegel  {\em et al.}~1998).

\subsection{         SDSS Colors of the Spaghetti Giants             }
\label{Spag}

Some regions observed by the Spaghetti survey overlap with runs 752
and 756. The SPS's Washington photometry is used to isolate metal-poor
stars on the basis of their $(M-T_2)$ and $(M-51)$ colors (sensitive
to temperature and the strength of the Mg$b$ and MgH features near
5200~\AA, respectively, cf Geisler et al. 1991; Paltoglou \& Bell
1994), and to obtain a first estimate of the luminosity class of the
stars. These candidates are observed spectroscopically, and are
classified into dwarfs and giants using the following indicators
(Morrison {\em et al.}~2002):
\begin{itemize}
\item the Mg$b$ and MgH features near 5200~\AA, which are characteristic
      of dwarfs, and are almost absent in giant stars for $0.8 \le (B-V)
      \le 1.3$ (Flynn \& Morrison 1990);
\item the CaI $\lambda$4227 line, which is usually present in dwarfs and
      absent in giants. While this feature may be visible in metal-poor
      giants (with [Fe/H]$\ge -1.5$), it is much weaker than in dwarfs
      of the same color;
\item the CaII H and K lines near 3950~\AA, which are sensitive to
[Fe/H] (cf Beers {\em et~al.}~1999).
\end{itemize}

The contours in the top left panel in Fig.~\ref{fig:spag_giants} show,
in the \gr\ vs. \ug\ color-color diagram, the distribution of 19,000
stars from run 125 that are brighter than $r=19$ and whose photometric
errors in all bands ($u$, $g$, $r$) are less than 0.05. Nine SPS
giants\footnote{Kindly provided to us by H. Morrison.}  (Dohm-Palmer
{\em et al.}~2001) are shown by filled circles and are clearly offset
from the center of the stellar locus (for a discussion of the position
of the stellar locus in the SDSS photometric system see e.g. Finlator {\em
et al.}~2000). In the other SDSS color-color projections the SPS
giants fall right on the stellar locus, having $0.15 \le \ri \le 0.4$
and $0 \le \iz \le 0.25$. Note also that since metal-poor stars are
bluer than metal-rich stars of the same temperature (cf Mihalas \&
Binney 1981), they are shifted left from the main locus in the \gr\
vs.  \ug\ diagram shown in Fig.~\ref{fig:spag_giants} (Lenz {\em et
al.}~1998; Fan 1999).

\subsection{         Definition of the $s$  Color                  }

We use the well-defined stellar locus to derive a principal axes
coordinate system (\Po, \Pt), where \Po\ lies parallel to the stellar
locus and \Pt\ measures the distance from it (see also Odenkirchen
{\em et al.}~2001; Willman {\em et al.}~2001).  The origin is chosen
to coincide with the highest stellar density (\ug$=1.21$,
\gr$=0.42$). Since the objects of interest occur in a relatively
narrow color range, we restrict ourselves to $1.1 \le \ug \le 2$ and
$0.3 \le \gr \le 0.8$ (we refer the reader to Lenz {\em et al.}~1998
for an extensive study on how the SDSS colors of stars translate into
temperature, metallicity and surface gravity). This procedure yields:
\begin{eqnarray}
              \Po =  0.910 \,(\ug) + 0.415 \,(\gr) -1.28, \\
              \Pt =  -0.415 \,(\ug) + 0.910 \,(\gr) + 0.12.
\end{eqnarray}

\begin{figure}
\plotone{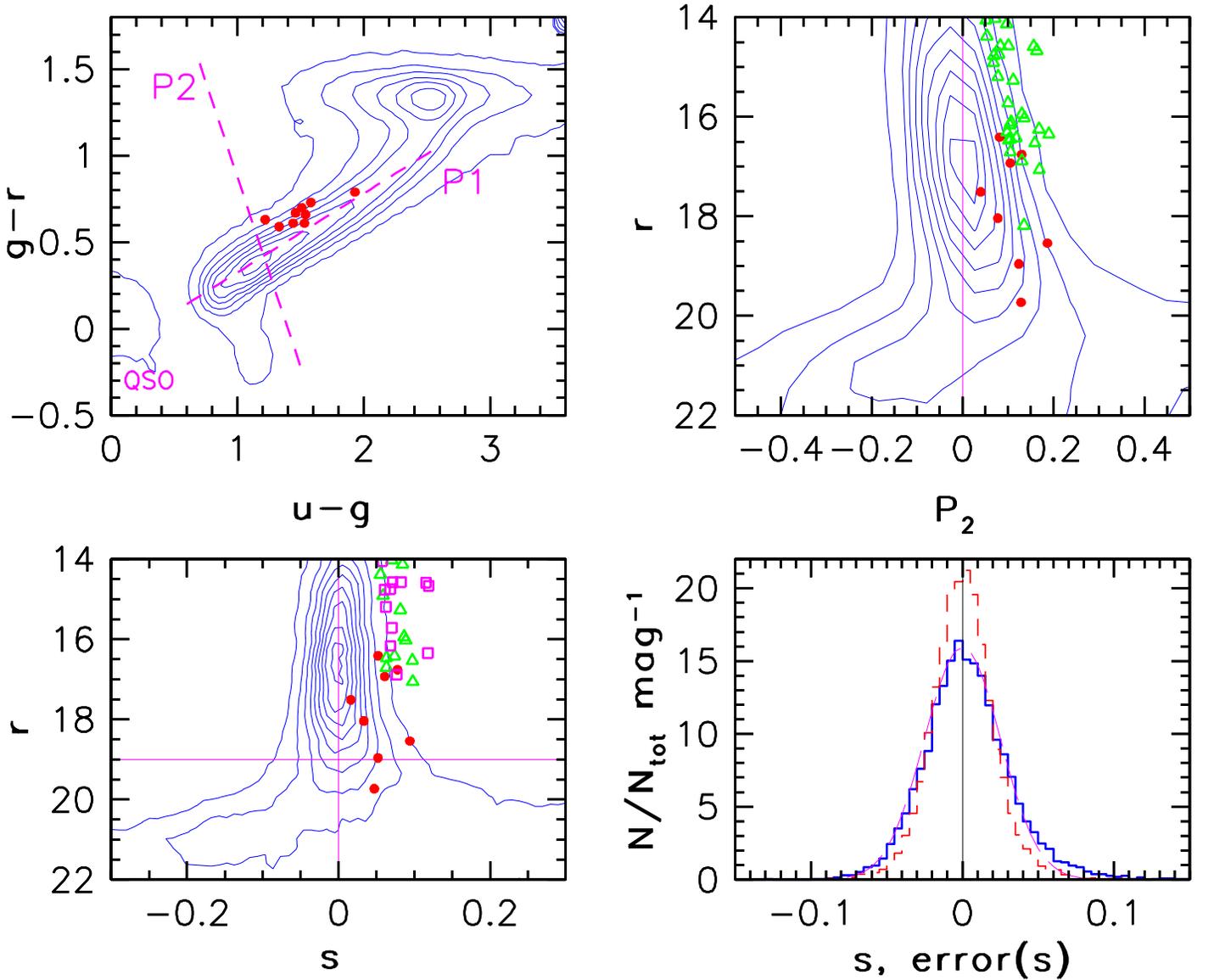}
\caption{The top left panel shows with contours the distribution of
19,000 stars with $r$$<$19 in the \gr\ vs. \ug\ color-color diagram. A
sample of metal-poor giants discovered by the Spaghetti survey is
shown by solid circles (all panels); note that they are offset from
the locus center. The dashed lines show a principal axes system
aligned with the locus.  The top right panel shows the $r$  vs. \Pt\
color-magnitude diagram, and the bottom left panel shows the $s$ 
vs. \Pt\ diagram, where $s$  is derived from \Pt\ by accounting for the
magnitude dependence.  The thick solid histogram in the bottom right panel
shows the distribution of the $s$  color for stars brighter than $r=19$
from run 125, that were also observed in run 1755. The dashed histogram
shows the $s$  error distribution determined from multiple
observations. The thin solid curve is a best Gaussian fit to the $s$ 
color distribution and shows that the latter is not symmetric: the red
wing has more stars than the blue wing. The candidate giants from the
spectroscopic sample discussed in Sec.~3.2 are shown by triangles
in the upper right panel, and divided into confirmed giants (squares)
and dwarfs (triangles) in the bottom left panel.
\label{fig:spag_giants}}
\end{figure}

The top right panel in Fig.~\ref{fig:spag_giants} shows the $r$ 
vs.~\Pt\ color-magnitude diagram for 47,771 stars from run 125.  The
position of the locus clearly depends on the $r$  magnitude (the median
\Pt\ color becomes redder at the faint end), and we correct for this
effect using a linear \Pt\ vs.~$r$  fit (the correction varies from
--0.03 to 0.05 mag). We thus define a new color,~$s$  -- named after
the Spaghetti survey--, that is normalized such that its error is
approximately equal to the mean photometric error in a single band
(assuming uncorrelated measurements in the $u$, $g$, and $r$  bands). We
obtain
\begin{equation}
     s =  -0.249 \, u + 0.794 \, g - 0.555 \, r + 0.24.
\end{equation}

The $r$  vs.~$s$  color-magnitude diagram for stars with $-0.1 < \Po <
0.6$ is shown in the bottom left panel in Fig.~\ref{fig:spag_giants}.
The thick solid line in the bottom right panel in
Fig.~\ref{fig:spag_giants} shows the distribution of the $s$  color for
stars brighter than $r=19$ that were observed both in runs 125 and
1755. The equivalent Gaussian distribution width determined from the
interquartile range is 0.035 mag.  The dashed line shows the
distribution of the difference in the $s$  color between the two epochs
divided by $\sqrt{2}$; its width is 0.025 mag; that is,
the error distribution is narrower than the observed $s$  color
distribution, demonstrating that the $s$  color distribution reflects
some intrinsic stellar property. The thin solid line is a best
Gaussian fit to the $s$  color distribution and shows that the latter
is not symmetric: the red wing contains more stars than the blue wing.

\subsection{    The Selection Criteria for Metal-poor Giants       }

Based on the $s$  color distribution of the Spaghetti giants and the
overall $s$  color distribution, we select candidate metal-poor giants
as stars with $r < 19$ and $-0.1 < \Po < 0.6$ (for $1.1 \le \ug \le 2$ and 
$0.3 \le \gr \le 0.8$) and 
\begin{equation}                    
\label{eqn:sel}
 s  > m_s + 0.05, 
\end{equation}
where $m_s$ is the median value of $s$ color in appropriately chosen
subsamples. Since the accuracy of EDR data calibration is finite
(about 0.01--0.02 mag), $m_s$ is not exactly zero. We calibrate
independently the data corresponding to a given run and camera column
(i.e. individual scanline), and compute $m_s$ for these subsamples.
As expected, the $m_s$ distribution is well described by a Gaussian
with a width of $\sim$0.025 mag (see top right panel of
Fig.~\ref{fig:752-756}).  Note as well that $m_s$ (or equivalently the
location of the principal axes) may change slightly as a function of
Galactic coordinates as a result of a different mixture of stellar
populations. However, this shift is sufficiently small that we may
neglect it. Hereafter, we will simply use $\hat{s}$ when referring to
this median-corrected color.

\section{              Tests of the Selection Method         }

\subsection{ Comparisons of the candidates and a control sample }

\begin{figure}
\plotone{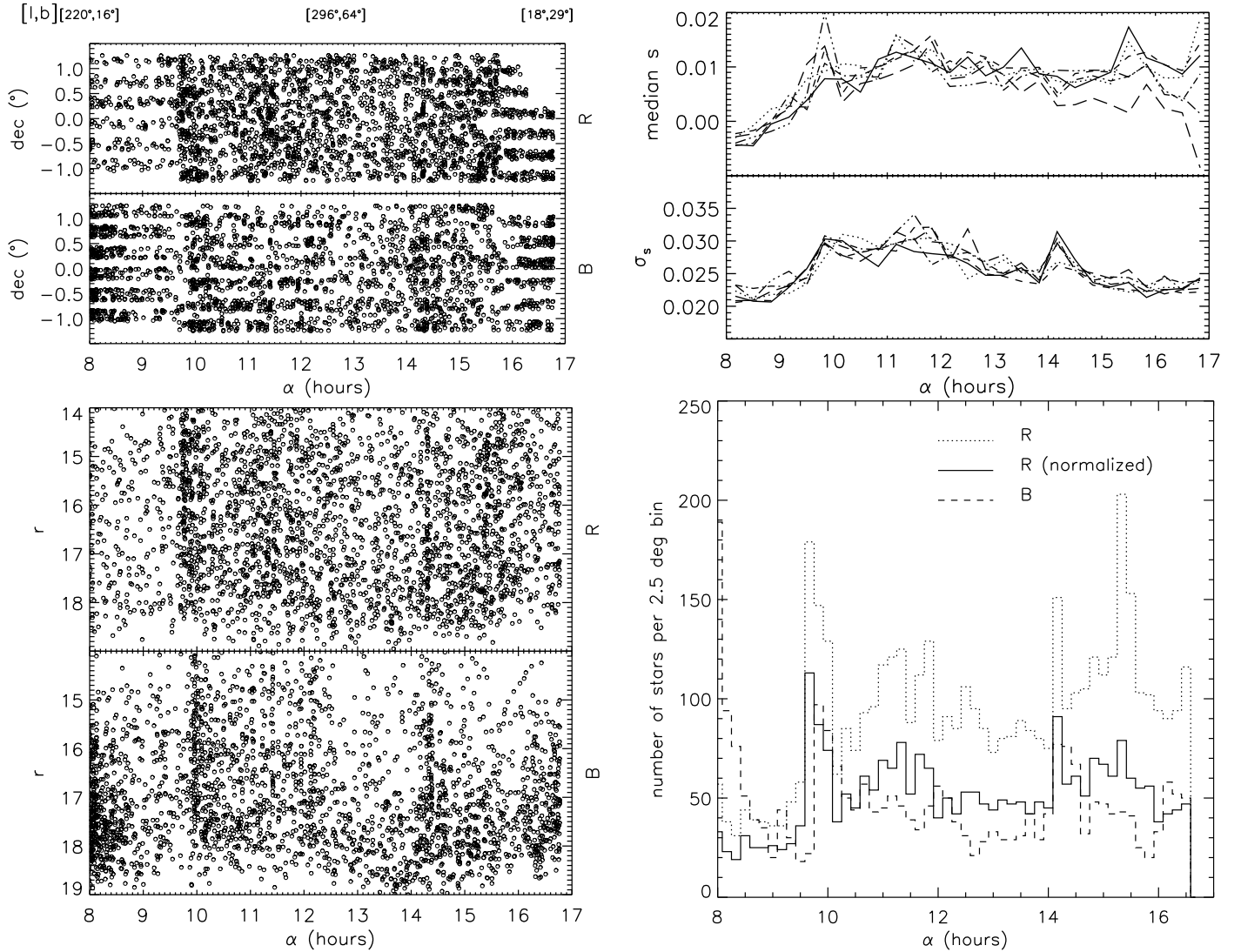}
\caption{The panels on the left show the sky distribution (declination
vs. right ascension $\alpha$) and $r$ magnitude vs. $\alpha$ diagrams
for stars in runs 752 and 756 satisfying Eq.(\ref{eqn:sel}).  There
are 5125 stars in \R\ and 2844 in \B\ . Here we have selected a random
sample of \R\ with the same number of stars as in \B\ for a more
direct comparison of the spatial distribution of stars in each
subset. The regions $\alpha \lesssim 10^h$ and $15.5^h \lesssim
\alpha$ contain data from a driftscan that does not have interleaving
stripes, hence the very uneven distribution of stars at different
declinations in those regions. The top right panels show the median
$s$-color and the dispersion as a function of $\alpha$ for the six
independent camera columns. We have excluded data closer than 150
pixels to the chip edge, because the flatfielding uncertainties in the
$u$ band increase photometric errors by about 1\%. The histograms in
the bottom right panel show the number of stars per 2.5 deg. bins for
the whole \R\ sample (dotted curve), for the \B\ sample (dashed curve)
and for a random realization of the \R\ sample with the same number of
stars as \B\ (solid curve).
\label{fig:752-756}}
\end{figure}

We define two samples for comparing the angular and magnitude
distribution: the ``candidates" \R\ with $\hat{s} > +0.05$, and the
``control sample" \B\ with $\hat{s} < -0.05$.  In all runs that we have
analyzed, we find that the number of stars in \R\ is significantly
larger than in \B. We compare the angular and magnitudes distribution
of the two samples in Fig.~\ref{fig:752-756} for stars in runs 752
and 756 whose photometric errors\footnote{Here we use a newer processing
rerun than available in EDR.} in all bands ($u$,$\,g$,$\,r$) are less
than 0.05.   To account for a smaller number of stars in \B\, we have
selected a random subset of \R\ having the same size as \B.

The top left panels in Fig.~\ref{fig:752-756} show that the angular
distribution of stars in \R\ appears more isotropic than that of
\B. In particular, the number of stars in \B\ increases toward lower
Galactic latitudes, indicating that they are dominated by the disk
population. It is evident from the bottom left panels in the same
Figure that the samples have different magnitude distributions -- the
\R\ sample contains a larger fraction of bright stars --, which
possibly reflects different distance distributions. These results hint
toward an overall different spatial distribution for stars in the
``candidate region" from those in the ``control sample". The bottom
right panel shows that there is an enhancement in the stellar density
of the \R\ sample (dotted curve) for $\alpha_{ 2000}$ in the range
$13^{h} - 16^{h}$, which can be linked to the recently discovered
clumps in the Galactic Halo, associated with the Sagittarius dwarf
northern tidal streams (Ivezi\'c {\em et al.}~2000; Yanny {\em et
al.}~2000; Dohm-Palmer {\em et al.}~2001; Martinez-Delgado {\em et
al.}~2001; Newberg {\em et al.}~2002). We performed a $\chi^2$
statistical test to determine the probability that the \B\ and \R\
samples are drawn from the same parent population. If we restrict the
samples to $\alpha_{ 2000} = 13^{h} - 16^{h}$, we find this probability
to be $\sim 3 \times 10^{-9}$. If no restriction is applied the
probability is even smaller $\sim 10^{-10}$.  The southern streams of
Sagittarius in runs 94 and 125 ($\alpha_{ 2000} \sim 1^h 15^m$,
$\delta_{ 2000} \sim 0^\circ$; Yanny {\em et al.}~2000), and the Draco
dwarf spheroidal galaxy in runs 1336/9 and 1356/9 ($\alpha_{ 2000} =
17^h 20^m$, $\delta_{ 2000} = 57.9^\circ$; Odenkirchen {\em et
al.}~2001), -- none of which is shown in Fig.~\ref{fig:752-756} -- are
recovered as clear overdensities in the distribution of ``candidate
stars", which is not the case for stars in the ``control sample".

\subsection{   Spectroscopic Observations and Analysis}

The results from the previous subsection suggest that the fraction of
halo giants in the candidate subset \R\ is significantly larger than
that of the comparison sample \B. To investigate this, and to derive
an estimate of the selection efficiency, we selected stars from two
overlapping runs 125 and 1755 which satisfy the selection criteria in
both runs. Of a total of 72 candidates with $\hat{s} > 0.05$, we
randomly selected 29, for which we obtained intermediate resolution
spectra.  The data reduced with the most recent version of photometric
pipeline ({\em photo} v5\_3) have photometric errors of the order 0.02
mag, smaller than the EDR data used here which have errors of the
order 0.03 mag.  The requirement that the candidate stars qualify in
both runs has a similar effect on efficiency as smaller photometric
errors, and thus the selection efficiency obtained here is more
representative of the upcoming SDSS Data Release~1.

The $r$ magnitudes of the selected candidates range from 14 to 17,
with a median $r$ magnitude of $16$ (see Table \ref{table1} for the
list of stars and their colors). The spectra were obtained during the
nights of October 20--24 2001, using CAFOS on the Calar Alto 2.2m
telescope in the framework of the ``Calar Alto Key Project for SDSS
Follow-up Observations'' (Grebel 2001).  The resolution was 4~\AA, the
spectral range $\lambda = 3200 - 5800$~\AA, and we integrated each
star for 900 up to 2000 seconds depending on the brightness of the
star and weather conditions. The reduction process will be described
in detail in Paper II.

\begin{figure}
\plotone{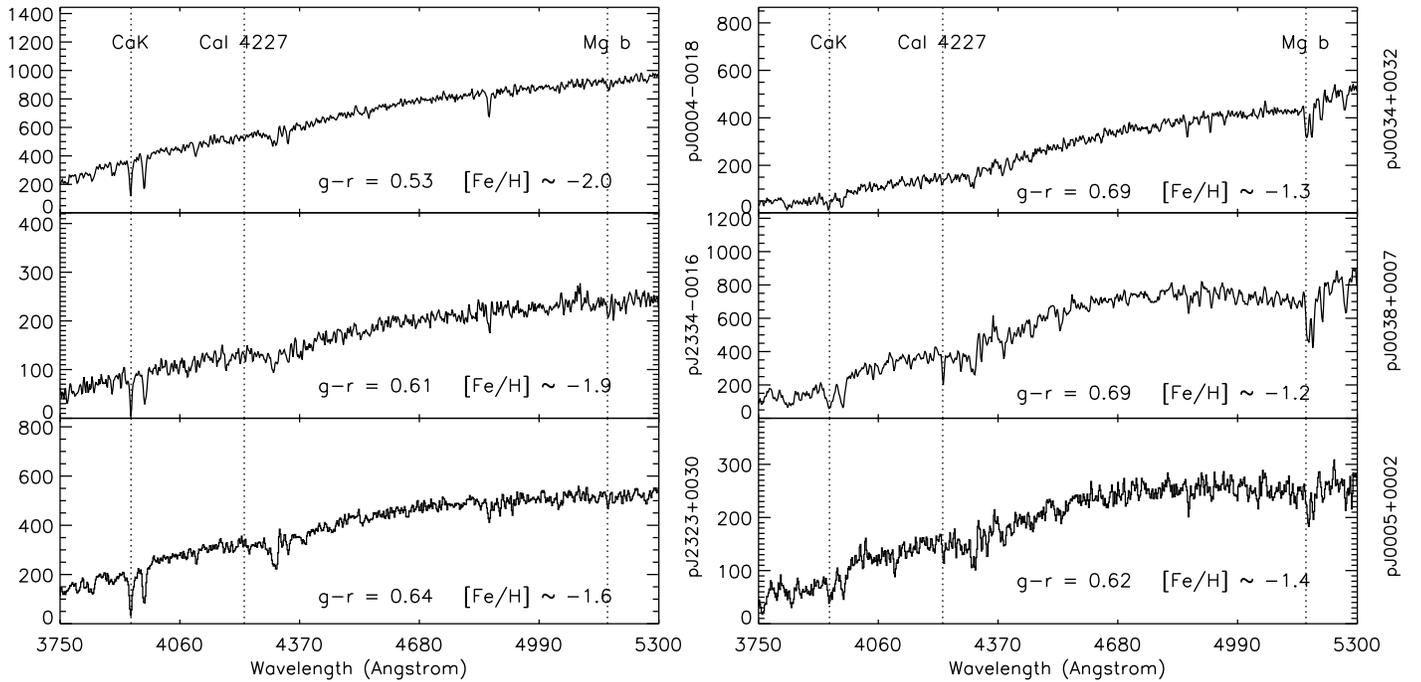}
\caption{Spectra of six ``candidate stars", which were observed with
the 2.2m telescope at Calar Alto. The spectra have 4~\AA\ resolution,
and have not been flux calibrated. The panels on the left show the
spectra of three giants, while those on the right correspond to three
dwarfs. Notice the strong Mg$H$ and Mg$b$ triplet in the dwarfs, which
are almost absent in the giants. The ``candidate stars" have
characteristic colors with median($g-r$) $=$ 0.62, which typically
corresponds to a $T_{eff} \sim 5000 K$, median($u-g$) $=$ 1.40, and
median($s$) $=$ 0.07. Their median $r$ magnitude is 16.03.
\label{fig:spectra}}
\end{figure}

In the wavelength range probed by the selection method, the features
most sensitive to luminosity class are those used by the SPS: the
Mg$b$ triplet near 5170\AA\ and the CaI $\lambda$4227 line.  In
Fig.~\ref{fig:spectra} we show the spectra of six of the stars in the
program. With the obtained resolution and signal-to-noise ratios, it
is possible to separate the giants from the dwarfs even by simple
visual inspection. The three spectra in the left panels clearly show
the absence of the MgH and Mg$b$ triplet (compare to the spectra shown
in the right panel), thus proving that at least some of the stars in
our sample are giants. To obtain a more quantitative discrimination
between giant and dwarf stars, we also measure the equivalent widths
of the CaII K, the CaI $\lambda$4227 and the Mg$b$ triplet lines,
according to the definitions by Morrison {\em et al.}~(2002). We use
their calibrations to assign luminosity class. Out of the 29 observed
stars, 12 are metal-poor giants, 13 are metal-poor dwarfs, and for 4
stars the obtained signal-to-noise ratio is insufficient for
classification.  A rough estimate of [Fe/H] has been obtained by
visual inspection of the spectra (we do not have enough standards for
proper calibration). Such a method has an inherent uncertainty of about 0.5
dex.

We conclude that the identification of metal-poor giants can be done
with satisfactory efficiency, of order 50\%, using the SDSS
photometric data. However, we caution that stars in our spectroscopic
sample are relatively bright ($r<17$), and thus this efficiency will
be lower than 50\% for fainter magnitudes due to increased photometric
errors and contamination by subdwarfs.

\section{                        Discussion                           }

The technique described in this Letter can be used to study the
structure of the Galactic halo in two complementary ways. One method
is to select candidates for spectroscopic follow-up and determine
their luminosity class and radial velocity. The latter essentially adds
an extra dimension which could prove useful to disentangle structures
in the halo (Harding {\em et al.}~2001). This approach would be
analogous to that used by the SPS. A second possibility, and which
makes perhaps a better use of the uniquely large SDSS dataset, is a
statistical approach.  One can compare the angular and magnitude
distributions of candidate giant stars with that of stars in a
``control sample", in the same spirit of Fig.~2. A statistical
subtraction of the distributions of the datasets may allow, for
example, a mapping of overdensities in the number of candidate giant
stars at different locations in the sky. When combined with other
techniques for selecting halo stars, such as RR Lyrae stars and blue
horizontal branch stars, it will be possible to produce an
unprecedented detailed three-dimensional map of the Galactic halo
based on the SDSS imaging survey.

\acknowledgments We are especially grateful to Heather Morrison and
the rest of the SPS collaboration for providing us with a list of
giants from the Spaghetti survey, and for very useful discussions.  
\v{Z}I and RHL acknowledge generous financial support and constant
encouragement by GRK, and \v{Z}I thanks the Max-Planck-Institute
for Astrophysics for hospitality. We
also thank Stefano Zibetti for the help in setting up IRAF at MPA.
The spectroscopic observations were made in the framework of the
``Calar Alto Key Project for SDSS Follow-up Observations'' at the
German-Spanish Astronomical Centre, Calar Alto Observatory, operated
by the Max Planck Institute for Astronomy, Heidelberg jointly with the
Spanish National Commission for Astronomy.  Funding
for the creation and distribution of the SDSS Archive has been
provided by the Alfred P. Sloan Foundation, the Participating
Institutions, the National Aeronautics and Space Administration, the
National Science Foundation, the U.S. Department of Energy, the
Japanese Monbukagakusho, and the Max Planck Society. The SDSS Web site
is http://www.sdss.org/.  The SDSS is managed by the Astrophysical
Research Consortium (ARC) for the Participating Institutions.  The
Participating Institutions are The University of Chicago, Fermilab,
the Institute for Advanced Study, the Japan Participation Group, The
Johns Hopkins University, Los Alamos National Laboratory, the
Max-Planck-Institute for Astronomy (MPIA), the Max-Planck-Institute
for Astrophysics (MPA), New Mexico State University, Princeton
University, the United States Naval Observatory, and the University of
Washington.

\begin{table}
\begin{center}
\title{List of spectroscopically observed giant candidates} 
\begin{tabular}{rrccccc}
\hline $\alpha_{2000} ~~(^h~ ^m~ ^s)$ & $\delta_{2000}~~(^\circ~ '~
'')$ & $u-g$ & $g-r$ & $r$ &$s$& ``class"\\

\hline
23 22  49.9 &       45.6 & 1.18 & 0.52 & 14.57 &  0.08 & G \\
23 23 30.5 &    58  17.9 & 1.47 & 0.75 & 16.35 &  0.12 & G \\
23 23 44.0 &    30  16.9 & 1.32 & 0.64 & 14.59 &  0.12 & G \\
23 25  51.1 & 1 $\,\,\,$  0  23.0 & 1.62 & 0.79 & 17.06 & 0.10 & D \\
23 28  57.0 &   55  49.8 & 1.56 & 0.50 & 16.31 &  0.06 & u \\
\smallskip
23 29  16.7 &   55  51.4 & 1.31 & 0.61 & 16.89 &  0.08 & G \\
23 31$\,$  $\,\,5.9$ &    2  21.7 & 1.47 & 0.66 & 15.27 & 0.08 & D \\
23 31  22.3 &   $-$17  32.4 & 1.33 & 0.59 & 16.45 & 0.07 & u \\
23 34  59.6 &   $-$16 $\,$   5.1 & 1.40 & 0.62 & 15.72 &  0.07 & G \\
23 37  35.1 &$-$1 $\,\,\,$  5 $\,$   2.5 & 1.71 & 0.77 & 16.68 & 0.07 & D \\
\smallskip
23 43  56.4 &      10  52.3 & 1.35 & 0.58 & 14.58 &  0.07 & G \\
23 49  39.7 &   $-$42  57.0 & 1.70 & 0.77 & 16.42 &  0.07 & D \\
23 51  53.9 &       2  37.9 & 1.60 & 0.71 & 16.71 &  0.06 & D \\
   4  56.7 &    $-$18  37.0 & 1.07 & 0.54 & 14.67 &  0.12 & G \\
   5$\,$  $\,\,7.3$ & 2  40.8 & 1.37 & 0.62 & 15.94 & 0.09 & D \\
\smallskip
  10  16.9 &       52  25.4 & 1.71 & 0.71 & 14.05 &  0.06 & G \\
  15  24.1 &    $-$21  26.9 & 1.38 & 0.69 & 16.25 &  0.11 & u \\
  21  57.5 &    $-$17  38.5 & 1.23 & 0.52 & 14.75 &  0.07 & G \\
  29$\,$  $\,\,0.8$ & 5 29.5 & 1.42 & 0.59 & 14.77 & 0.06 & G \\
  30$\,$  $\,\,5.8$ & $-$24 $\,$ 7.2 & 1.62 & 0.78 & 16.53 & 0.10 & D \\
\smallskip
  32  20.1 &    51  41.4 & 1.09 & 0.48 & 16.17 & 0.07 & G \\
  32  31.5 & $-$22  27.2 & 1.61 & 0.72 & 16.47 & 0.06 & D \\
  33  36.4 & $-$49  $\,$  8.7 & 1.42 & 0.57 & 14.38 & 0.06 & D \\
  33  40.5 & $-$17  17.5 & 1.42 & 0.62 & 14.13 & 0.08 & D \\
  34  28.4 &    32  $\,$  4.3 & 1.47 & 0.69 & 16.03 & 0.09 & D \\
\smallskip
  35  44.5 &     4  12.4 & 1.30 & 0.54 & 15.19 & 0.06 & G \\
  38  21.8 & $-$17  48.4 & 1.35 & 0.56 & 14.02 & 0.07 & D \\
  38  39.2 &     7  12.9 & 1.64 & 0.69 & 14.91 & 0.06 & D \\
  38  42.4 &    35  55.1 & 1.23 & 0.55 & 16.11  & 0.07 & u \\
\hline
\end{tabular}
\end{center}
\caption{Right ascension, declination, apparent magnitudes and colors
of the 29 candidates with follow$-$up spectra. The last column
denotes whether the star is a giant (G), dwarf (D) or whether the
spectra was not good enough to determine the luminosity class (u).}
\label{table1}
\end{table}


\vspace*{-0.5cm}
\begin{thebibliography}{}
\bibitem[Beers et al.(1999)]{1999AJ....117..981B} Beers, T.~C., Rossi, S.,
Norris, J.~E., Ryan, S.~G., \& Shefler, T.\ 1999, \aj, 117, 981
\bibitem[Dohm-Palmer {\em et al.}(2001)]{rdp01} Dohm-Palmer, R.C., {\em et
    al.}~2001, \apj, 555, L37
\bibitem[Fan(1999)]{1999AJ....117.2528F} Fan, X.\ 1999, \aj, 117, 2528
\bibitem[Finlator(2000)]{Finlator}Finlator, K., {\it et al.}~2000, AJ, 120, 2615
\bibitem[Flynn \& Morrison(1990)]{1990AJ....100.1181F} Flynn, C.~\&
Morrison, H.~L.\ 1990, \aj, 100, 1181
\bibitem[Fukugita et al.(1996)]{1996AJ....111.1748F} Fukugita, M., 
Ichikawa, T., Gunn, J.~E., Doi, M., Shimasaku, K., \& Schneider, D.~P.\ 
1996, \aj, 111, 1748 
\bibitem[Geisler 1984]{geisler} Geisler, D., 1984, PASP, 96, 723
\bibitem[Geisler, Claria, \& Minniti(1991)]{1991AJ....102.1836G} Geisler,
D., Claria, J.~J., \& Minniti, D.\ 1991, \aj, 102, 1836
\bibitem[gr]{gr} Grebel, E. K., 2001, Reviews in Modern Astronomy, 14, 223
\bibitem[Gunn]{Gunn}Gunn, J.E., {\it et al.}~1998, AJ, 116, 3040
\bibitem[Harding et al.(2001)]{2001AJ....122.1397H} Harding, P., Morrison,
H.~L., Olszewski, E.~W., Arabadjis, J., Mateo, M., Dohm-Palmer, R.~C.,
Freeman, K.~C., \& Norris, J.~E.\ 2001, \aj, 122, 1397
\bibitem[Helmi et al. 2002]{h02} Helmi, A., White, S.D.M., \& Springel,
V., 2002, MNRAS in press, astro-ph/0208041
\bibitem[Hogg]{Hogg}Hogg, D.W., Finkbeiner, D.W., Schlegel, D.W., \&
            Gunn, J.E. 2002, AJ, 122, 2129
\bibitem[Ivezic et al 2000]{ive00} Ivezi\'{c}, \v{Z}. {\em et al.} 2000,
\aj, 120, 963
\bibitem[Johnston et al. 1996]{kvj96} Johnston, K.V., Hernquist, L., \&
Bolte, M., 1996, ApJ, 465, 278
\bibitem[Kinman, Suntzeff, \& Kraft(1994)]{1994AJ....108.1722K} Kinman,
T.~D., Suntzeff, N.~B., \& Kraft, R.~P.\ 1994, \aj, 108, 1722
\bibitem[Lenz]{Lenz} Lenz, D. D., Newberg, H. J., Rosner, R., Richards,
G. T., \& Stoughton, C. 1998, ApJS, 119, 121
\bibitem[Lupton]{Lupton}Lupton, R.H. {\it et al.}~2001, in {\it Astronomical
Data Analysis Software and Systems X}, ASP Conf. Proc.,
Vol.238, p. 269. eds. F.R. Harnden, Jr., F. A. Primini, \&
H. E. Payne
\bibitem[Majewski, Ostheimer, Kunkel, \&
Patterson(2000)]{2000AJ....120.2550M} Majewski, S.~R., Ostheimer,
J.~C., Kunkel, W.~E., \& Patterson, R.~J.\ 2000, \aj, 120, 2550
\bibitem[Martinez-Delgado, Aparicio, G{\' o}mez-Flechoso, \& 
Carrera(2001)]{2001ApJ...549L.199M} Martinez-Delgado, D., Aparicio, 
A., G{\' o}mez-Flechoso, M.A., \& Carrera, R.\ 2001, \apjl, 549, 
L199 
\bibitem[Mayer et al. 2002]{mayer02} Mayer, L., Moore, B., Quinn, T.,
Governato, F., \& Stadel, J., 2002, MNRAS, 336, 119
\bibitem[Mihalas \& Binney (1981)]{mb} Mihalas, D. \& Binney J.,
1981, ``Galactic Astronomy'', Freeman, p.~120
\bibitem[Morrison et al.(2000)]{hlm00} Morrison, H.L., {\em et al.}
    2000, \aj, 119, 2254
\bibitem[Morrison et al.(2001)]{2001AJ....121..283M} Morrison, H.~L.,
Olszewski, E.~W., Mateo, M., Norris, J.~E., Harding, P., Dohm-Palmer,
R.~C., \& Freeman, K.~C.\ 2001, \aj, 121, 283
\bibitem[Morrison et al.(2002)]{hlm02} Morrison, H.L., {\em et al.}
2002, submitted to AJ
\bibitem[Newberg et al.~2002]{2002ApJ...569..245N} Newberg, H.~J., Yanny, B., 
Rockosi, C., {\em et al.}, 2002, \apj,  569, 245 
\bibitem[Odenkirchen et al.(2001)]{2001AJ....122.2538O} Odenkirchen,
M.~{\em et al.}, 2001, \aj, 122, 2538 
\bibitem[Paltoglou \& Bell(1994)]{1994MNRAS.268..793P} Paltoglou, G.~\&
Bell, R.~A.\ 1994, \mnras, 268, 793
\bibitem[Pier {\em et al.}~2002]{astrom} Pier, J.R., {\em et al.}
         2002, AJ, in press
\bibitem[Robin et al.~2000]{2000A&A...359..103R} Robin, A.~C., Reyl{\' e}, 
C., Cr{\' e}z{\' e}, M., 2000, \aap,  359, 103 
\bibitem[Schlegel, Finkbeiner, \& Davis(1998)]{1998ApJ...500..525S}
Schlegel, D.~J., Finkbeiner, D.~P., \& Davis, M.\ 1998, \apj, 500, 525
\bibitem[Smith]{Smith}Smith, J.A., {\em et al.}~2002, AJ, 123, 2121
\bibitem[Sommer-Larsen, Flynn, \& Christensen(1994)]{1994MNRAS.271...94S}
Sommer-Larsen, J., Flynn, C., \& Christensen, P.~R.\ 1994, \mnras, 271, 94
\bibitem[Steinmetz \& Navarro(2002)]{2002NewA....7..155S} Steinmetz, M.~\& 
Navarro, J.~F.\ 2002, New Astronomy, 7, 155 
\bibitem[Stoughton et al.(2002)]{2002AJ....123..485S} Stoughton, C.~{\em et al.}\
2002, \aj, 123, 485
\bibitem[Vivas et al.(2001)]{2001ApJ...554L..33V} Vivas, A.~K.~{\em et al.}\
2001, \apjl, 554, L33
\bibitem[Yanny et al 2000]{yan00} Yanny, B. {\em et al.}~2000, \apj,
540, 825
\bibitem[york]{york}York, D.G. {\em et al.} 2000, AJ, 120, 1579
\bibitem[Willman et al.~2002]{2002AJ....123..848W} Willman, B., Dalcanton, 
J., Ivezi{\' c}, {\v Z}., Jackson, T., Lupton, R., Brinkmann, J., Henessy,
G., Hindsley, R., 2002, \aj,  123, 848 
\end{thebibliography}
\end{document}